**Dispersion of Boron Nitride Powders in Aqueous Suspensions with Cellulose**

Florian Bouville[1], Sylvain Deville[1]

*[1]Laboratoire de Synthèse et Fonctionnalisation des Céramiques, UMR3080 CNRS/Saint-Gobain, Cavaillon, France*

**Abstract**

The dispersion of nitride and carbide ceramic particles in water is difficult, due to the absence of hydroxyl groups on their surface. Boron nitride is not an exception and despite its numerous applications, no effective dispersant has been identified so far. We demonstrate here the dispersion properties of two cellulose derivatives for hexagonal boron nitride powders, hydroxyl ethyl cellulose and methyl cellulose. The effect of particle sizes and cellulose concentration was investigated. The adsorption of cellulose onto the surface of the BN particles was confirmed by isotherm adsorption. Zeta potential measurements shows a charge screening effect of the cellulose The suspensions obtained were highly loaded and stable versus pH, and thus could lead to homogeneous codispersion of BN particles and sintering additives.

**I. Introduction**

Hexagonal boron nitride (BN) exhibits a unique combination of high thermal conductivity, high resistance to thermal shock, high thermal/chemical stability, and low surface energy. These properties arise from its crystalline structure[1]: covalently bonded boron and nitrogen that forms layers that are weakly bonded by Van der Walls forces. The structure is similar to graphite but electrically insulating. Because of its low surface energy, BN is particularly useful in applications involving liquid metals handling where it acts as a barrier to prevent corrosion and contamination. Such benefits are also particularly useful for high temperature crucibles[3]. These two main applications usually rely on aqueous powder processing routes, to cast a preform or deposit a coating. The low sinterability of BN requires then the use of densification techniques such as hot isostatic pressing (HIP) in presence of sintering additives[4,5].

Unfortunately, the low surface energy of BN is a major drawback for its use in aqueous media[6], because it is not easily dispersed. This dispersion issue often results in heterogeneities in the green body, both in terms of particles packing density and sintering additives. An effective dispersant could allow the use of highly loaded suspensions with a homogenous mix of BN particles and sintering additives. The final ceramic parts could therefore present more homogenous microstructures and thus improved mechanical properties.

Cellulose derivatives are usually added in suspension as thickening agent, to prevent the settling of large particles. We demonstrate through zeta potential, viscosity, and adsorption measurements that cellulose are also efficient dispersants of BN particles in aqueous suspensions.

## II. Experimental procedure

BN powders were provided by Saint-Gobain BN products (Très BN series). Those powders are mainly used in cosmetics products and thus present a relatively high purity and crystallinity, but also low boric acid content. The main characteristics of the different powders are given in the table 1. The granulometric distributions have been obtained by laser granulometry, which assumes a relative sphericity of the particles. The equivalent diameters measured are the diameter of the platelets, values confirmed by SEM observations. The thicknesses of the platelet have been evaluated by SEM observations.

Two types of cellulose were tested: a hydroxyethyl cellulose HEC (Tylose H 4000 P2 from ShinEtsu, $M_v$ = 720 000) and methyl cellulose MC (Methocel A4M from Dow Chemicals, $M_w$ = 88 000). The celluloses present the same viscosity of 4 Pa.s at 2 wt.% in water but their structures are different, the MC is more hydrophile than the HEC. The suspensions were prepared by mixing the powder, the cellulose, and distilled water with a magnetic stirrer or a propeller mixer depending on their viscosity. Suspensions with hydroxyethyl cellulose were mixed for at least 4 hours to ensure total hydration before being used. Ultrasonic mixing was applied with a sonotrode (Digital Sonifier 250) with an applied energy around 150 W.h/kg.

Adsorption isotherm measurements were conducted using a total organic carbon (TOC) analyzer (TOC VCSN, Shimadzu), which provides a quantitative measure of the non-adsorbed fraction of cellulose in solution. BN/Cellulose suspensions from 0 to 1 wt.% of HEC were prepared. After complete hydration and mixing, the suspensions were centrifuged at 7200 rpm for 1h and supernatant aliquots were taken to be analyzed.

Zeta potentials were measured with a ZetaProbe after dispersion of the particles in distilled water by sonication. The pH was adjusted with 1 mol/L hydrochloric acid and sodium hydroxide. To calculate the zeta potential with this technique, the relative sphericity of the particles is assumed. This assumption is not valid with the anisotropic particles used here, so the measured values offer only a qualitative comparison of the zeta potential values between samples.Rheological measurements were performed in a concentric cylinder viscosimeter (Bohlin, Malvern, UK). The suspension was pre-sheared at 100 $s^{-1}$ during 30 s followed by 30 s at rest. Two procedures were followed: a shear rate ramp was applied from 0.05 $s^{-1}$ to 600 $s^{-1}$ or a constant shear rate for 1 min. A shear rate of 150 $s^{-1}$ was chosen to compare the viscosity of the different suspensions. Most of the suspensions have reached a quasi-Newtonian behavior before this value, which ensure the stability of the measure.

### III. Results and Discussion

#### (1) Effect of ageing time

A 19 vol.% of BN-16 suspension, with HEC to powder ratio of 0.019, was mixed by a magnetic stirrer at 800 rpm and a part of the suspensions was taken at different ageing times to perform rheological measurements. The evolution of viscosity versus ageing time is plotted in figure 1. The viscosity decreased by a factor six just by maintaining the stirring for fifteen days. The final viscosity was even slightly below the viscosity of the solvent with the same amount of HEC (0.11 Pa.s for the slurry and 0.15 Pa.s for 1 wt.% HEC solution). The evolution could be explained by the adsorption of the cellulose on particle surface. This phenomenon could help the breakage of agglomerates, and thus could reduce the suspension viscosity.

**(2) Effect of amount of cellulose**

Rheological characterization (fig. 2) has been conducted on 19 vol.% BN-8 and BN-1 suspensions with HEC to powder weight ratio ranging from 0 to 0.03. Concerning the BN-1 suspensions, only the HEC/BN weight ratios of 0.01, 0.02, and 0.03 were tested because the suspensions corresponding to the other ratios presented a viscosity too high to be assessed by the concentric cylinder rheological setup.

Repeated HEC additions minimized the viscosity of the suspensions for both particle sizes. After this minimum, the viscosity increases at different rates for the various particle sizes. The optimum weight ratio of HEC/BN is 0.0025 for 8 µm powder and 0.02 for 1 µm powder. The presence of a viscosity minimum is generally associated with a ratio where the dispersant molecules cover the maximum of particle surface without any excess in the suspension [7]. If we assume that HEC acts as a dispersant for BN, this particular ratio must be constant with the evolution of surface area. Suspension viscosity evolution versus HEC/powder ratio divided by powder surface area is shown in figure 2b. The minimum viscosity is reached at almost the same ratio for both powders, which is consistent with an optimum repartition of a dispersant on particle surface, independent of its size.

**(3) Adsorption isotherm of cellulose**

The adsorption isotherm for HEC on BN-8 is plotted in the figure 3. A nearly total adsorption on the surface (>90% for 10 mg/m²) is observed for a relative low amount of cellulose added (1g/L corresponding to a HEC/powder ratio of 0.0025). Then the adsorbed amount reached a plateau at 15 mg/m² when more cellulose is added. Those value are relatively high, compared to the typical value of well-known system such as poly(acrylic acid) (PAA) on oxides, that are around 1 mg/m² [8,9]. This quantity is usually related to the polymer conformation on particle surface, the more the polymer is stretched, the lower the concentration needed to entirely cover the surface. However, the cellulose used here presents chains nearly an order of magnitude longer (the molecular mass is $M_V$ = 720 000) than the usual dispersant used ($M_W$ around 10 000 for most of the commercially available PAA). This amount could come from the larger number of cellulose chains needed to cover the surface. The large molecular weight could also explain why a full coverage is not achieved. The cellulose derivatives are effectively adsorbed on boron nitride platelet and thus play the role of a steric dispersant.

**(4) Effect of cellulose on zeta potential**

Due to the low boric acid content (<0.1 wt.%), zeta potential of these BN particles is low, between -8 mv to -20 mV. Only the boric acid present on particle surface exhibit hydroxyl groups and thus can induce surfaces charges changes with the pH[10]. This behavior has also been reported by Crimp et al.[6] with two different powders of different boric acid content in aqueous suspension. The higher the surface charge, the higher the oxygen content. However, the addition of the HEC decreased the zeta potential from -20 mV to -4 mV at the solution pH and at pH 4. The decrease in zeta potential is usually associated with a more agglomerated state of particles, which induces higher suspension viscosity, but HEC addition lowered the viscosity (*figure 4*). The addition of this type of cellulose possibly masks the surface charge, thus lowering the zeta potential, and acts as a steric barrier to prevent agglomeration. This behavior of polysaccharides has been observed previously by Wang and al[11] with clay and ethyl (hydroxylethyl) cellulose. Clay particles have a flake morphology (a few microns in diameter and a hundred of nanometer thick) with a flat surface usually presenting a low negative surface charge (with a small numbers of hydroxyls groups), while the edges are positively charged (with hydroxyl group). Even if the interactions between cellulose and minerals are not well established yet, the frequently accepted explanation consists in adsorption due to a mix of hydrophobic forces on faces and hydrogen bonding on edges. Wang *et al.*[11] conclude that hydrogen bonding –thus interactions with edges– is predominant. With BN, edges should not play a key role in interaction due to a different particle size and aspect ratio. A similarity between clay faces and BN can be noticed because none of them present hydroxyl groups, and thus supposedly the main mechanism of cellulose adsorption on surface is hydrophobic forces. The zeta potential is lowered in neutral to acidic pH range, a range where cellulose is considered stable; a proper dispersion can theoretically be obtained within this range.

**(5) Effect of cellulose type on dispersion**

Another type of cellulose, methyl cellulose (MC), has been mixed with BN and the viscosity measured under the same conditions. MC was chosen to assess if the hydrophilicity influences the dispersion behavior. The resulting viscosity at different ratio and the data obtained with HEC are represented in figure 5.

An optimum viscosity is found for both MC and HEC at different cellulose/powder ratio (0.0025 and 0.005 for HEC and MC respectively). This suggests that the nature and/or amount of cellulose functional groups do not change its capacity to disperse the BN particles. Indeed, if the main mechanism of adsorption is hydrophobic bonding, the nature of cellulose does not have great impact as long as they have the same type of interaction with the solvent and particle, e.g. the same type of hydrophobic and hydrophilic functional groups. The amount of HEC required is slightly lower, possibly because of a larger molecular mass ($M_w$ = 80 000 for MC and $M_v$ = 720 000 for HEC) that leaded to a maximum coverage of the surface with less molecules.

**(6) Effect of solid loading**

The effect of solid loading on the behavior of a suspension is a critical parameter in order to use the slurry in typical ceramic processing, like slip or tape casting. A high aspect ratio of particles is a drawback to obtain loaded suspension because of low packing ability due to a greater steric hindrance. This can be partially tackled if the particle-particle interactions are limited due to an optimum amount of dispersant. In order to as determine maximum solid loading of cellulose-dispersed BN, the rheological behavior at different solid loading have been investigated (figure 6), using the dispersant /BN ratio obtained previously (figure 2).

A typical shear thinning behavior is obtained for the three different slurries, due to the breaking of particles arrangement and their alignment along the flow direction as the shear gradient increases[12].

Experimental data are plotted in figure 6 for the BN suspensions. For comparison a suspension of BN-8 without HEC at 17 vol.% was also investigated. The suspension without HEC presents a high shear-thinning behavior, its viscosity decreases by two orders of magnitude. The viscosity at low shear rate of the suspension with HEC is more than an order of magnitude lower than the viscosity of the suspension without HEC. The viscosities values of both suspensions get closer as a higher shear rate is applied, probably because the agglomerates are broken by the applied stress. The high viscosity at low shear rate makes it hard to cast the suspension in any mold. The dependencies of the relative viscosity (viscosity of suspension at high shear rates on viscosity of the medium[9]) values with the solid volume fraction $\phi$ are plotted in figure 7 with and without HEC. The relative viscosities without dispersant are higher at lower volume fractions (three times higher for a volume fraction of 0.3). The data with dispersant were fitted by a Krieger–Dougherty model[9,10] (1). The high shear-thinning behavior of the suspensions without cellulose additives makes it hard to precisely determine the viscosity minimum and thus to apply the same model on it.

$$\eta_{rel} = (1 - \frac{\phi}{\phi_m})^{-\phi_m [\eta]} \qquad (1)$$

Equation (1) allows calculating $\Phi_m$, the volume fraction of particle where the viscosity tends to the infinity. In this equation, $\eta$s is the viscosity of the dispersing medium and $[\eta]$ is the intrinsic viscosity. The intrisic viscosity and the maximum volume fraction are fonction of particle's shape. The value of $[\eta]$ = 2.5 and $\Phi_m$ = 0.64 are obtained with perfectly spherical and monodispersed particle, but $[\eta]$ increases and $\Phi_m$ decreases as the particle aspect ratio increases. The value obtained with BN, $[\eta]$ = 3.1 and $\Phi_m$ = 0.36, are consistent with these evolutions and the values found in litterature with various materials[13,14]. Barnes *et al.*[12] found an empirical formula for the intrinsic viscosity evolution with the aspect ratio of disks $[\eta]$ = 3 (axial ratio) /10 and leads, in this case, to an aspect ratio of approximately 10, which is consistent the aspect ratio of 16 for BN particles, considering the high polydispersity of these powders (cf. table 1). As described previously, the plate-like shape of these particles prevents a high solid loading suspension.

**IV. Conclusion**

Hexagonal boron nitride (BN) aqueous slurries can be dispersed using cellulose derivatives. The mechanism of the adsorption on particles surface is not explained in this study although experimental evidences point toward hydrophobic forces. Because cellulose is relatively stable versus pH, the dispersion can be achieved within a relatively large pH range. Obtaining a higher volume fraction could be achieved by adjusting more precisely the molecular weight.

A stable dispersion with a high solid loading offers interesting possibilities to obtain homogeneous green bodies with fewer defects, especially when associated with electrostatic dispersion of sintering additives. An alternative to make BN ceramic can now be considered: the use of thermal gel casting. Indeed cellulose ethers form a gel when heated above a certain temperature and thus could directly be used for ceramic shaping.

**Acknowledgements**

We acknowledge the financial support of the ANRT (Association Nationale Recherche Technologie) and Saint-Gobain through a CIFRE fellowship, convention #808/2010.

**References**

1. Ooi, N., Rairkar, A., Lindsley, L. & Adams, J. B. Electronic structure and bonding in hexagonal boron nitride. Journal of Physics Condensed Matter 18, 97–115 (2006).

2. Lipp, A., Schwetz, K. A. & Hunold, K. Hexagonal boron nitride: Fabrication, properties and applications. Journal of the European Ceramic Society 5, 3–9 (1989).

3. Eichler, J. & Lesniak, C. Boron nitride (BN) and BN composites for high-temperature applications. Journal of the European Ceramic Society 28, 1105–1109 (2008).

4. Ertug, B., Boyraz, T. & Addemir, O. Microstructural aspects of the hot-pressed hexagonal boron nitride ceramics with limited content of boron oxide. Materials Science Forum 554, 197–200 (2007).

5. Lee, S.-K., Nakamura, K., Kume, S. & Watari, K. Thermal conductivity of hot-pressed BN ceramics. Materials Science Forum 510-511, 398–401 (2006).

6. Crimp, M. J., Oppermann, D. A. & Krehbiel, K. Suspension properties of hexagonal BN powders : effect of pH and oxygen content. Journal of Materials Science 4, 2621–2625 (1999).

7. Briscoe, B. J., Khan, A. U. & Luckham, P. F. Optimising the dispersion on an alumina suspension using commercial polyvalent electrolyte dispersants. Journal of the European Ceramic Society 18, 2141–2147 (1998).

8. Cesarano, J., Aksay, I. A. & Bleier, A. Stability of Aqueous alpha-Al2O3 Suspensions with Poly(methacrylic acid) Polyelectrolyte. Journal of the American Ceramic Society 71, 250–255 (1988).

9. Kirby, G. H., Harris, D. J., Li, Q. & Lewis, J. A. Poly(acrylic acid)-Poly(ethylene oxide) Comb Polymer Effects on BaTiO 3 Nanoparticle Suspension Stability. Journal of the American Ceramic Society 87, 181–186 (2004).

10. Lewis, J. a. Colloidal processing of ceramics. Journal of the American Ceramic Society 83, 2341–2359 (2000).

11. Wang, J. & Somasundaran, P. Mechanisms of ethyl(hydroxyethyl) cellulose-solid interaction: influence of hydrophobic modification. Journal of colloid and interface science 293, 322–32 (2006).

12. Barnes, H. A., Hutton, J. F. & Walters, K. An introduction to rheology. 201 (1989).

13. Bergstrom, L. Shear thinning and shear thickening of concentrated ceramic suspensions. Colloids and Surfaces A: Physicochemical and Engineering Aspects 133, 1998 (1998).

14. Mora, M., Gimeno, F., Amaveda, H., Angurel, L. a. & Moreno, R. Dispersant-free colloidal fabrication of $Bi_2Sr_2CaCu_2O_8$ superconducting thick films. Journal of the European Ceramic Society 30, 917–926 (2010).

**Figures**

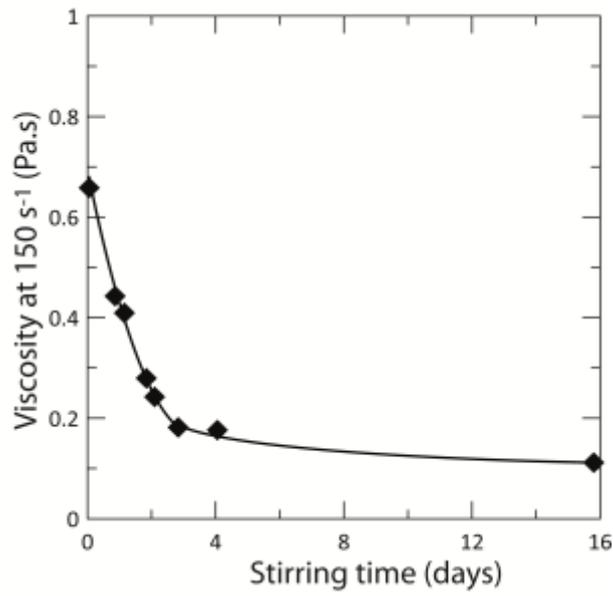

Fig. 1. Viscosity vs the ageing time of 19 vol.% BN-16 suspension.

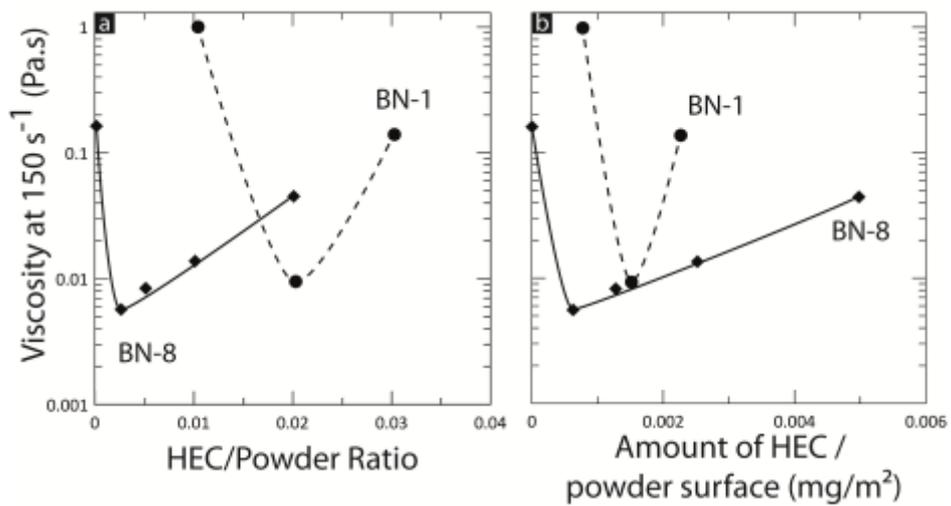

Fig. 2. (a) Effect of HEC/Powder weight ratio on different sizes of boron nitride particles. (b) Effect on viscosity of HEC/powder ratio divided by powder specific area.

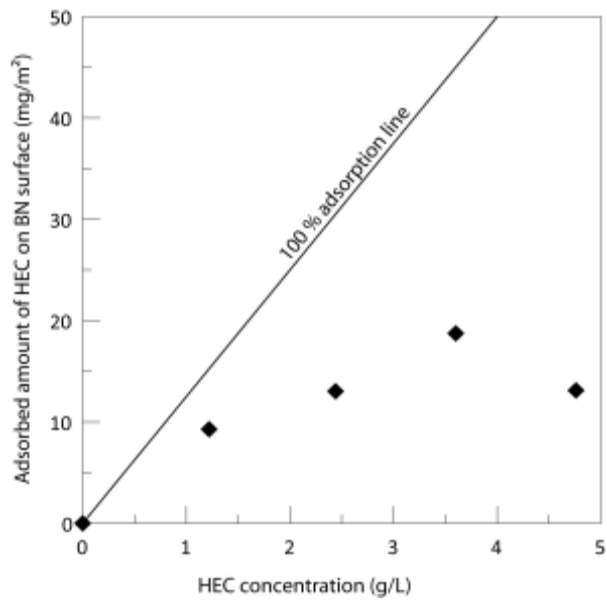

Figure 3. Adsorption isotherm of HEC on BN-8 surface.

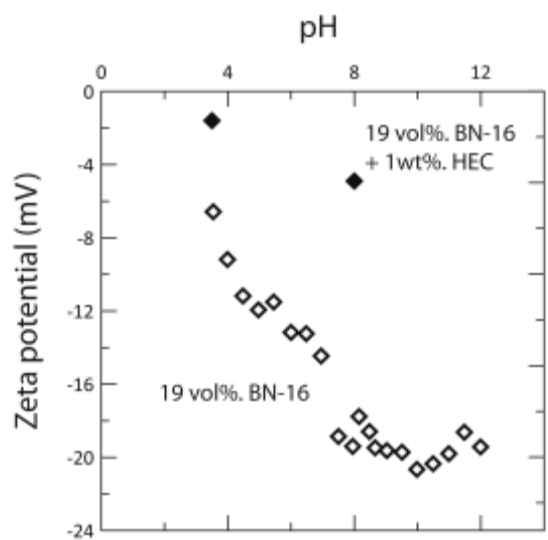

Fig. 4. Effect of HEC on Zeta potential BN particles.

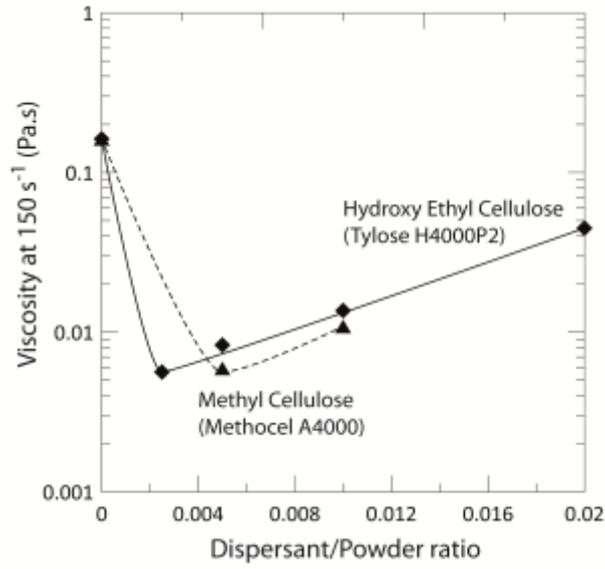

Fig. 5. Effect of cellulose type on suspension viscosity.

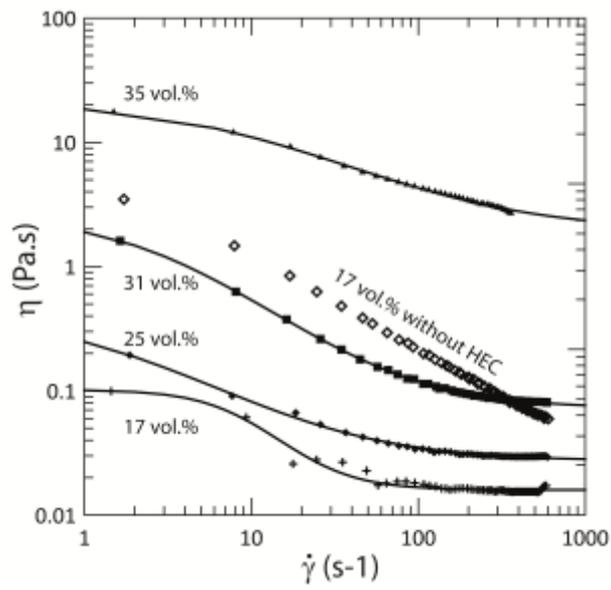

Fig. 6. Rheological behavior of BN-8 with and without HEC suspension at different solid loadings.

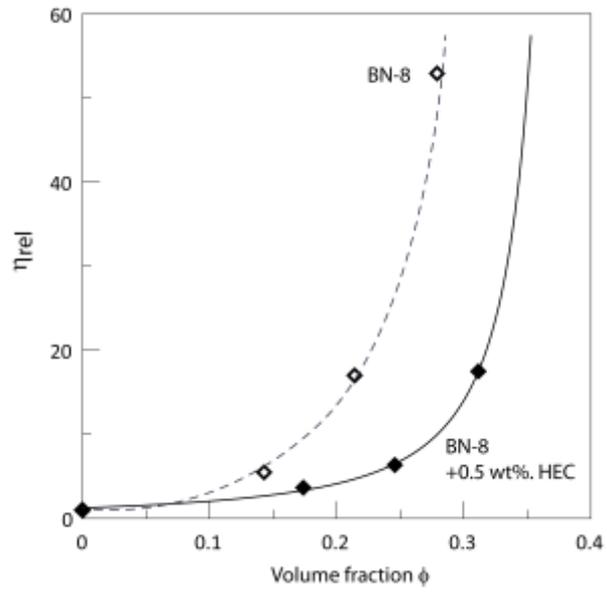

Fig. 7. Minimum viscosity versus particle volume fraction with and without HEC. The solid line is the Krieger-Dougherty model fit with $\Phi = 0.36$ and $[\eta] = 3.1$ parameters value.